\documentclass[12pt,a4paper,english]{article}
\usepackage{amssymb,amsmath,graphicx,multicol}
\pdfoutput=1
\usepackage[T1]{fontenc}
\usepackage{subfigure}
\usepackage{graphicx}
\usepackage{amssymb}
\usepackage{esint}
\usepackage{a4wide}
\usepackage{psfrag}   
\usepackage{babel} 
\usepackage{epstopdf}
\newcommand{\be}{\begin{equation}} 
\newcommand{\ee}{\end{equation}}
\newcommand{\bea}{\begin{eqnarray}}
\newcommand{\eea}{\end{eqnarray}}
\newcommand{\bi}{\begin{itemize}}
\newcommand{\ei}{\end{itemize}} 
\newcommand{\bc}{\begin{column}{0.50\textwidth}} 
\newcommand{\ec}{\end{column}}
\newcommand{\bcs}{\begin{columns}} 
\newcommand{\ecs}{\end{columns}}
\makeatletter

\@addtoreset{equation}{section}

\@addtoreset{figure}{section}

\@addtoreset{table}{section}
\makeatother

\begin{document}
\begin{flushright}
{\normalsize SISSA 04/2010EP}
\end{flushright}

\vspace*{0.5cm}
\begin{center}
\LARGE
Energy Level Distribution of Perturbed \\
Conformal Field Theories
\end{center}
\vspace{0.5cm}
\begin{center}
{\large G. P. Brandino$^{1,2}$, R.M. Konik$^{3}$ and 
G.\ Mussardo$^{1,2,4}$
\vspace{0.9cm}}

{\sl $^1$International School for Advanced Studies (SISSA),\\
Via Beirut 2-4, 34014 Trieste, Italy\\[2mm]
$^2$INFN, Sezione di Trieste\\[2mm]
$^3$ Condensed Matter and Material Science Department, Brookhaven National Laboratories \\ Upton, NY USA\\[2mm]
$^4$The Abdus Salam International Centre\\ for Theoretical Physics, Trieste, Italy
}

\end{center}

\vspace{0.5cm}
\begin{center}
{\bf Abstract}\\[5mm]
\end{center}
We study the energy level spacing of perturbed conformal minimal models in finite volume, considering perturbations of such models that are massive but not necessarily integrable. We compute their spectrum using a renormalization group improved truncated conformal spectrum approach. With this method we are able to study systems where more than 40000 states are kept and where we determine the energies of the lowest several thousand eigenstates with high accuracy. We find, as expected, that the level spacing statistics of integrable perturbed minimal models are Poissonian while the statistics of non-integrable perturbations are GOE-like.  However by varying the system size (and so controlling the positioning of the theory between its IR and UV limits) one can induce crossovers between the two statistical distributions.

\newpage

\section{Introduction}

The distinction between integrable and non-integrable models is a central problem in many fields of physics and mathematics. The question is important in the quantum case, particularly for those systems with infinite degrees of freedom and local interactions. In this case, the only general result available consists in the identification of one-dimensional systems as the only models where integrability survives to the quantum level \cite{Coleman}. This no-go theorem, while limiting quantum integrability to one-dimensional theories, has, at the same time, led to intense focus on these same systems, both in their lattice and continuum formulations, leading to a number of remarkable developments in many-body physics (see, for instance \cite{ZZ,BPZ,Zamo1,Baxter,Mattis,Sutherland,Mussardo,EssKon}).
The precise control offered by integrable field theories allows one to also examine non-integrable models, both in the massive \cite{DMS} or massless cases \cite{CM},
through perturbation theory based on the exact matrix elements of the perturbed operator. In such cases the nature of the non-integrable theory depends crucially on whether the perturbation is local or non-local with respect to the excitations of the original integrable model. In the former case, the mass of the excitations moves proportionally to the strength of the perturbation and therefore the excitations below the lowest threshold remain stable while those above threshold acquire an imaginary part in their mass and then decay \cite{DGM}.  In the latter case, much more dramatic results occur. There the non-local nature of the perturbation gives rise to confinement phenomena of the excitations, with a consequent drastic change in the spectrum 
\cite{DMS,CM,McCoy-Wu} (see also \cite{LMT,MT,DGconf,FZsp,Rut,Coldea}). 

Despite an impressive mass of knowledge collected over the years following the Coleman-Mandula theorem, it is fair to say that we are still missing a well-established set of theoretical tools for studying in a satisfactory way several aspects of non-integrable systems. Lacking, for instance, is a correspondent to the KAM theorem \cite{kam} that in classical mechanics describes the effects of the breaking of an integrable Hamiltonian and gives the conditions under which quasi-periodic motion survives. Also lacking, significantly, is a proper understanding of time evolution of quantum systems with infinite degrees of freedom.  The importance of this issue and its intertwining with the system's integrability or non-integrability is seen in the series of experiments on one-dimensional cold atom systems.  Depending on the dynamics and the boundary conditions to which these have been subjected, the experiments have produced two sets of results.  In one, the time evolution of the systems is insensitive to the initial conditions with a consequent thermalization of the final state \cite{Hofferberth}. In the second \cite{Kinoshita}, the system has shown no sign of thermalization, i.e. the systems did not equilibrate even after thousands of collisions of the particles.  

The aim of this paper is to demonstrate a further discriminator between integrable and non-integrable quantum systems. Obviously, a general strategy in identifying integrable models is to show that the model admits a number of non-trivial conserved quantities equal to the number of degrees of freedom.  This number becomes infinite in the case of a quantum field theory. Following this route is a difficult task for many models. There are, however, other features that provide hints of whether a given quantum system is integrable or not.  The structure of energy level spacings in a quantum model is one such feature \cite{Berry}. In particular the presence of an infinite number of conserved charges should give rise to a high degeneracy in the energy levels, while a non-integrable system should present no sign of such degeneracies, provided we take into account "trivial" symmetries such as translational invariance, discrete symmetries, and so on.

In the light of these remarks, we focus our attention here on the spectrum of a special class of quantum Hamiltonians obtained by deformations of the minimal unitary models of conformal field theories (CFT) \cite{BPZ}.  The minimal unitary models, whose Hamiltonian is denoted by the sequence $H_p$ ($p = 3,4,\cdots$), are the continuum limit of certain (often elementary) quantum lattice models.  The simplest of the sequence, $H_3$, corresponds to the Ising model at its critical point.  The next in the sequence, $H_4$, is the tricritical-critical Ising model. $H_5$ is the three-state Potts model, and so on.  Alternatively, the sequence corresponds to a set of Landau-Ginzburg theories where the potential of the order parameter field, $\Phi$, is a polynomial of order $\Phi^{2(p-1)}$ \cite{ZamLG}.
Or equally good, the quantum Hamiltonians $H_p$ are those which enter the transfer matrix of the two-dimensional RSOS lattice statistical models with maximal height 
$p+1$ \cite{Andrews}.

We focus on these theories for several reasons. First and foremost the minimal models and their perturbations offer a rich variety exhibiting all the different desired classes
of behaviour. CFT's are themselves integrable field theories and for the minimal model subclass, everything can be explicitly calculated.  Furthermore perturbations of minimal CFTs can be both integrable \cite{Zamo1} and non-integrable. Secondly, CFT's and their deformations, irrespective of whether they are integrable or not, admit a Hamiltonian formulation in a geometry of an infinite cylinder of circumference $R$. 
As shown by Al. Zamolodchikov \cite{Zamo2}, the matrix elements of the Hamiltonian in this geometry can be obtained using conformal data alone. 

This formulation allows one to numerically study such theories through what is known as 
the truncated conformal space approach (TCSA).  The ``truncated'' in the TCSA refers to the need to truncate the Hilbert space at some level (even in finite volume, the minimal models are defined on a continuum space and so their corresponding Hilbert space is infinite dimensional). While all aspects of the perturbed minimal
models can be accessed in this fashion (despite the truncation) -- including the
masses of the lowest excitations, matrix elements, and even decay rates of unstable particles \cite{Takacs} -- we focus here on the energy levels.  Because we are interested
in level spacing statistics (rather than, more typically, the particular value of one low-lying energy level), we need to accurately compute (at least) on the order of 4000-5000 levels of the system. In order to do this in the context of the TCSA, we employ a numerical renormalization group adapted to the particulars of the TCSA \cite{Konik}.  The numerical renormalization group used here is akin to the Wilsonian numerical renormalization group used in the study of quantum impurity problems.
It is then a variational approach where the full Hamiltonian is approximated by a matrix
product ansatz.  To ensure that the best possible ansatz is employed, we introduce here
a sweeping procedure. This ensures the thousands of needed energy levels are accurately determined.

The paper is organized as follows.  In Section 2, with the help of some simple examples, we remind the reader of the main motivations to study a model's level statistics.  We also state the result of a general theorem that a set of uncorrelated variables has level spacing statistics governed by a (generalized) Poisson distribution. In Section 3 we present both how the TCSA works and the necessary details of the conformal minimal models relevant to its implementation. In particular we describe how the numerical renormalization group is implemented with the TCSA and how we have expanded the procedure to include sweeps to optimize results. In Section 4 we present our results on the level spacing statistics for a number of integrable and non-integrable models.  We also demonstrate that by varying the system size $R$ (at fixed perturbation strength)
we can induce a crossover from the Poissonian statistics applicable to the level spacings of an integrable model to the level spacings of a non-integrable model. This crossover occurs generically because at sufficiently small $R$, the perturbation is weak and the
level spacing statistics is that of the unperturbed model, while at sufficiently large $R$ the truncation effects of the TCSA render any perturbation non-integrable.

\section{Level spacing statistics in non-integrable and integrable models}

Michael Berry originated the idea of studying level spacing statistics to analyze integrable and non-integrable behaviour \cite{Berry}. In particular he was interested in the question of how a quantum system (which is in a sense deterministic) can exhibit chaotic motion in the classical limit. In this respect, a quantum integrable system was conjectured to exhibit deterministic motion in this limit, while in this same limit
a non-integrable quantum system sees chaotic motion.  The spectrum of the families of quantum systems was then expected to show distinct features, in particular regarding the distribution of level spacings.

Berry conjectured that the level spacing of an integrable system is Poissonian:
\be 
g(s)\,=\,e^{-s} \,\,\,,
\ee 
while the spacing of a non-integrable system 
is a Wigner (or Gaussian Orthonormal Ensemble, GOE) distribution:
\be 
g(s)\, =\, \frac{\pi s}{2} e^{-\frac{\pi}{4} s^2}\,\,\,,
\ee
where $s$ is the {\it normalized} spacing.  If $\{E\}_{i=1}^N$ is the 
set of energy levels of the systems, the normalized spacing corresponding to the i-th level is $s_i = (E_i-E_{i-1})\rho(E_i)$ where $\rho(E_i)$ is the density of states at energy $E_i$ \cite{Berry2,Poilblanc,Kudo}. This normalization ensures the mean spacing is 1.

While Berry himself provided good arguments for the integrable case \cite{Berry}, to the best of our knowledge, there is no general proof of this conjecture 
although it has been supported by several subsequent studies (see, for instance, 
\cite{Berry,Poilblanc,Brody,McDonald,Camarda,Seligman,Grosche,Rigol2}).  
While dealing with any particular system, care needs to taken in calculating the spacing distribution.  One must work 
in the subspaces where all the quantum numbers of the trivial symmetries that a system might have 
(e.g. translational invariance, discrete symmetries, spin reversal) have been fixed.
In a recent paper \cite{Kudo} an unexpected spacing distribution was found for the spin zero sector of a non-integrable XXZ chain. The spacing distribution resembled an average between a Poissonian and a GOE. Later the authors realized \cite{Kudo2} 
that a spin reversal symmetry, present in the spin zero sector alone, was not taken into account. Taking into account such degeneracies resulted in a GOE distribution as expected. 

\subsection{Some basic considerations and examples}

Before considering the level spacing of perturbed minimal models, we will examine a toy model so as to gain some intuition. Imagine a quantum Hamiltonian depending upon a parameter $l$.  Plotting the energy levels vs. $l$, we should be able to see 
levels crossing one another if symmetries are present, while we should see instead avoided crossings in the absence of all symmetry. The reason behind this is that in a given symmetry subspace, states are coupled together by the Hamiltonian, while in
different subspaces, states are decoupled and so evolve independently as we vary $l$.

\begin{figure}[ht]
\centering
\includegraphics[width=0.5\textwidth]{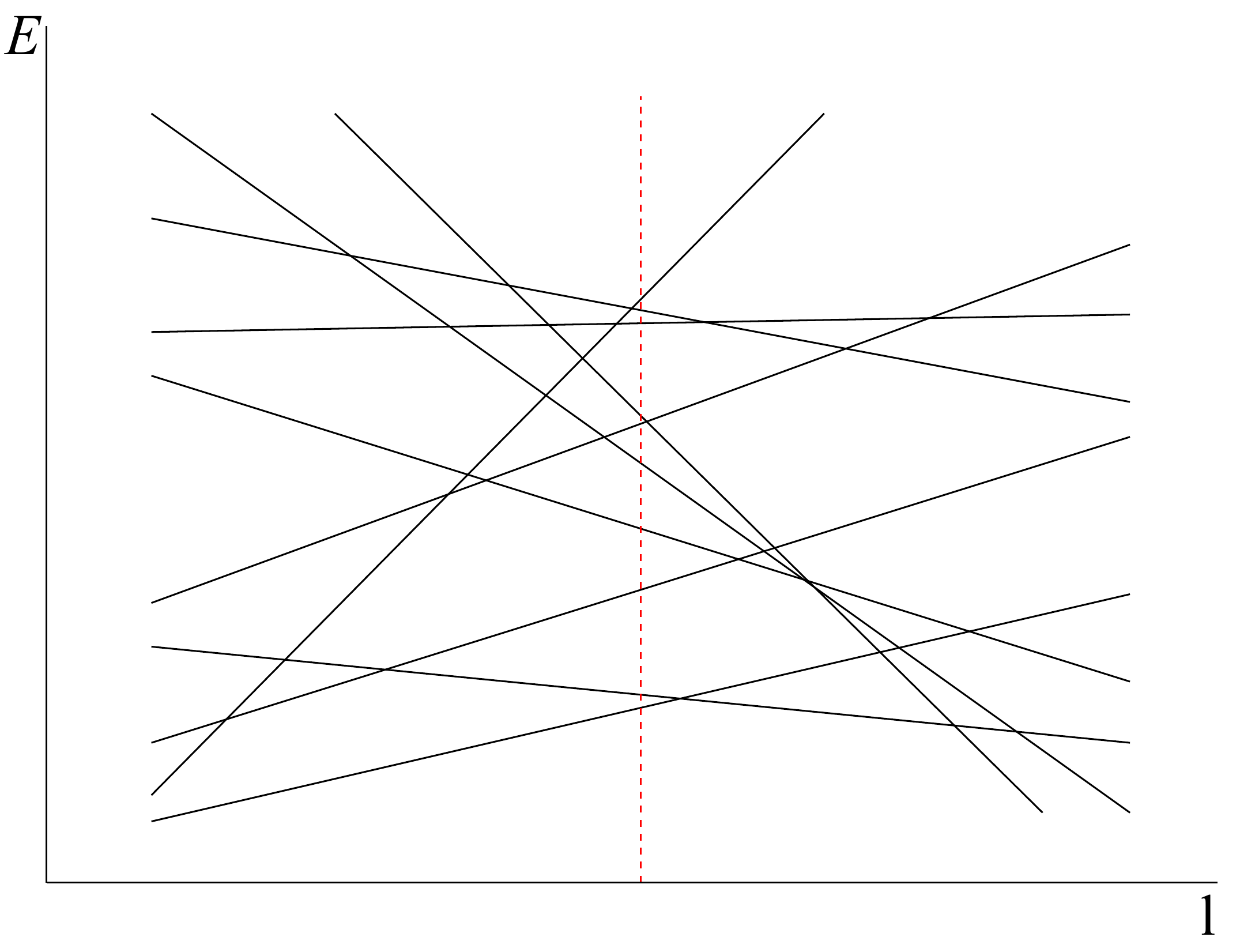}
\label{random-lines-es}
\caption{A set of lines with random slopes and intercepts. 
The spacing distribution is evaluated for a specific value of $l$, indicated by the red line in the figure.}
\end{figure} 

A naive way to model such a situation is to consider a set of straight lines on the plane, whose slope and intercept are independently identically distributed (i.i.d.) random numbers with a flat probability density (see Figure \ref{random-lines-es}).  
Consider the y-axis as the energy and fix a certain value of $l$.  The values of the ordinates of the lines at this $l$ then are interpreted as a set of energy levels.  The level spacing distribution of such a set of energy levels is shown in Figure 2.2.
\begin{figure}[ht]
\centering
\includegraphics[width=0.5\textwidth]{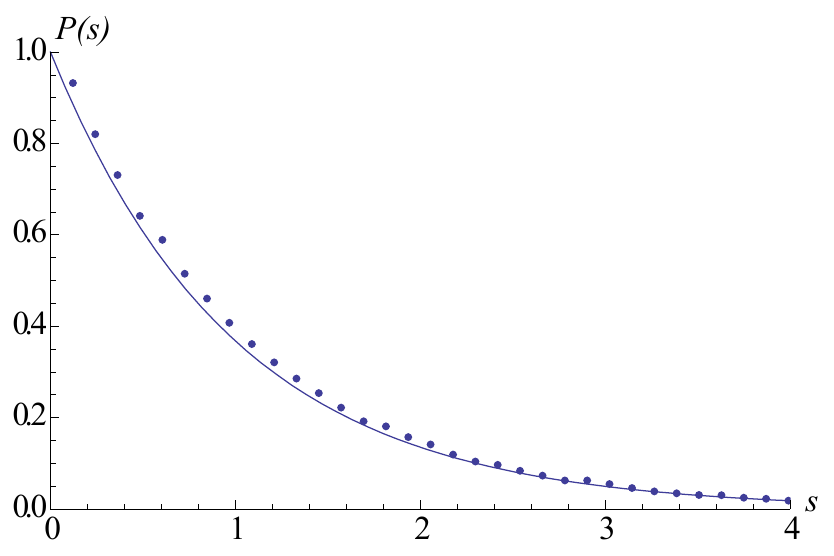}
\label{random-lines}
\caption{Spacing distribution for 100 000 random lines on a plane. The continuous line is the Poissonian distribution.}
\end{figure}
The distribution is clearly a Poissonian. This result could have been guessed from a remarkable theorem in probability theory by Pyke \cite{Pyke}. He showed that if we consider i.i.d. random numbers with probability density $f(x)$, the limiting level 
spacing distribution is given by
\bea
g(s)=\int_S f^2(x)e^{-sf(x)} dx; \\ 
\int_S f(x) dx=1 .
\eea
where $S$ is the support of $f(x)$. For $f(x)=\text{const}$ this results in a Poissonian distribution. It is interesting to notice that it is impossible to obtain a $g(s)$ that tends to zero for $s=0$ (like the GOE) since this would imply $\int_S f^2(x)dx=0$, a condition clearly impossible. This is by no means unexpected. As we argued before, the levels in a non-integrable system in a given symmetry sector are correlated and cannot be mimicked by independent variables. On the other hand, the levels of an integrable 
system are split into an infinity of distinct independent subspaces and so behave largely as uncorrelated variables.

\begin{figure}[b]
\begin{center}
$\begin{array}{ccc}
\includegraphics[width=0.33\textwidth]{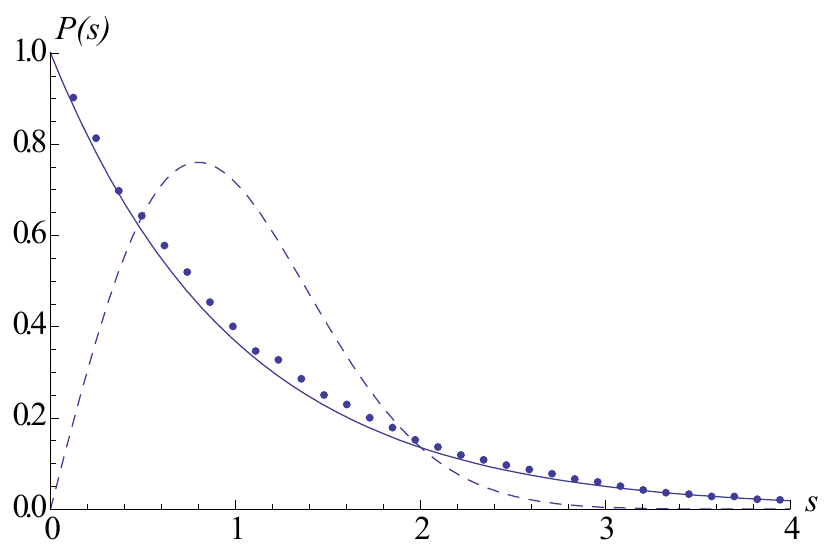}  
&\includegraphics[width=0.33\textwidth]{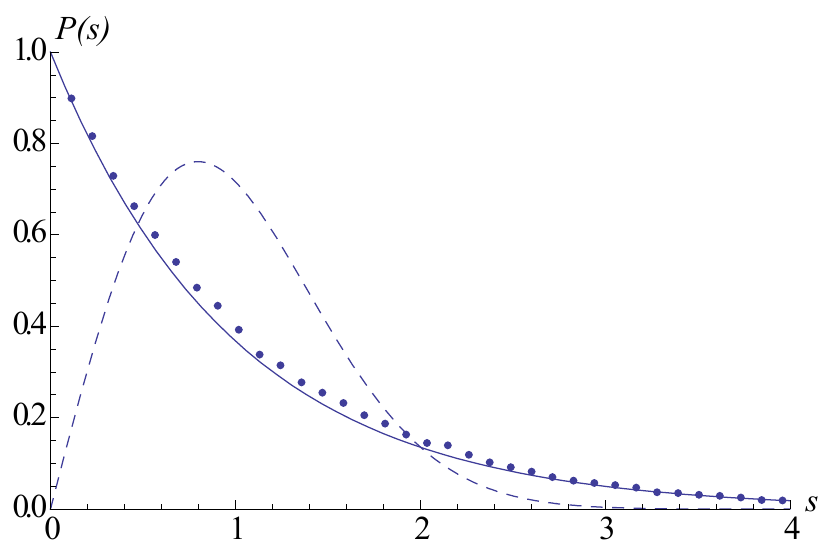} 
& \includegraphics[width=0.33\textwidth]{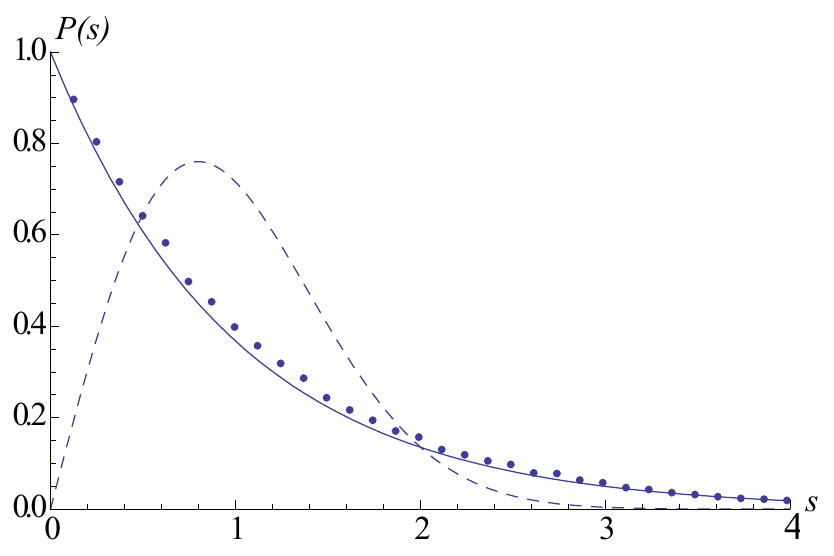} \\
(a) & (b) & (c) 
\end{array}$
\end{center}
\caption{Level spacing distribution for Gaussian correlated random variables with 
100 000 samples taken.
(a) r=0.1; (b) r=0.5; (c) r=0.9. For comparison we present the Poissonian distribution as a continuous
line while the GOE is given by a  dashed line.}
\label{gaussian}
\end{figure}

We may think that the reason why a distribution with $g(0)=0$ is impossible to get from i.i.d. random variables is their independence. To test such a statement we generated a set of correlated Gaussian random variables, with mean value $\mu=0$, 
variance $\sigma^2=1$ and correlation $r$. The trick to generate such a sample is straightforward. Start with $N+1$ independent Gaussian random variables $x_1,..,x_N,y$ with $\mu=0$ and $\sigma^2=1$. Define the linear combinations  
\begin{eqnarray}
z_i &=& x_i\sqrt{1-r}+y\sqrt{r} \label{defin1}; \cr
\zeta &=& \sum_{i}^N x_i \frac{\sqrt{r}}{\sqrt{1+r(N-1)}} - y\frac{\sqrt{1-r}}{\sqrt{1+r(N-1)}},
\label{defin2}
\end{eqnarray}
with $0\leq r\leq 1$. Since these are linear combinations, the $z_i$, $\zeta$ are still Gaussian random variables. However they are now correlated. Indeed, from definitions (\ref{defin1}) 
\begin{eqnarray}
\langle z_i\rangle &=& 0; \cr
\langle z_i z_j \rangle &=& r, ~~~~ i\neq j; \cr
\langle z_i z_j \rangle &=& 1, ~~~~ i=j;   \cr
\langle z_i \zeta \rangle &=& 0; \cr
\langle \zeta^2 \rangle &=& 1,
\end{eqnarray}
as required. 
The variable $\zeta$ is independent. Focusing on $z_i$, we obtain $N$ Gaussian correlated variables with the required properties. The spacing distribution for a sample of 100 000 of such variables is presented in Figure \ref{gaussian} for several values of $r$. Independent of this r, the distribution is still Poissonian.

\begin{figure}[ht]
\begin{center}
\includegraphics[width=0.5\textwidth]{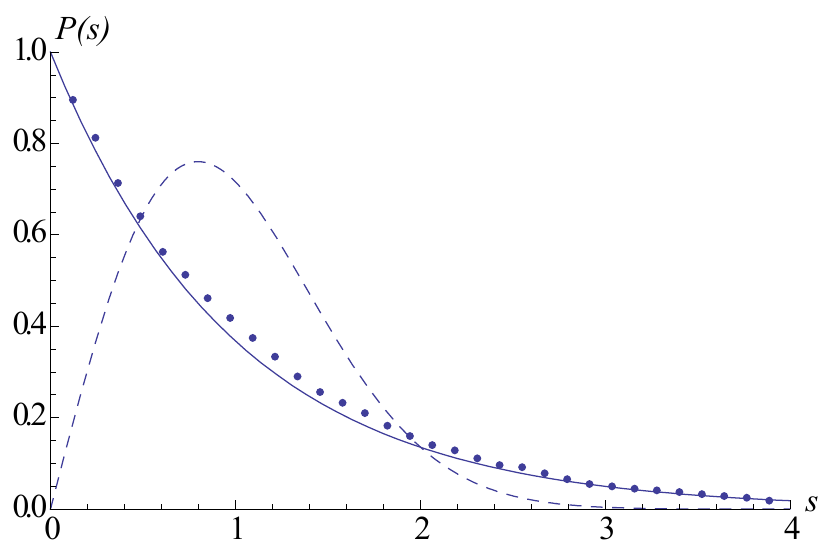} 
\end{center}
\caption{Level spacing distribution for non-Gaussian correlated random variables with 100 000 samples taken and r=0.5. Continuous line: Poissonian distribution; dashed line: GOE. } 
\label{nongaussian} 
\end{figure}

One may question whether a linear transformation is sufficiently complex enough to alter the spacing distribution. 
We thus try instead a quadratic transformation
\be
z_i=x_i^2\sqrt{\frac{1-r}{2}}+y^2\frac{-\sqrt{2(1-r)}+2\sqrt{4r-1} }{6},
\ee
with $\frac{1}{4}\leq r\leq 1$. The new variables are no longer Gaussian, but have correlation $r$ and variance 1.
But again the level spacing distribution is Poissonian (see Figure \ref{nongaussian}).

We conclude from this exercise that the correlation between variables required to get a GOE-like curve is much more involved. In random matrix theory the starting point is a random Hamiltonian $H$, symmetric for the GOE case, and so having $\frac{N(N+1)}{2}$ 
independent random variables.  The joint probability distribution of the $N$ eigenvalues corresponding to this $H$ is given by
\be
P(E_1,....E_N)\propto \prod_{j<k}(E_j-E_k)e^{-\frac{1}{2}\sum_{i=1}^N E_i^2}.
\ee
The term $\prod_{j<k}(E_j-E_k)$ is the responsible for the peculiar level spacing correlation.  It comes from the Jacobian of the transformation from the original matrix elements $H_{ij}$ to the energies $E_i$ (for details see \cite{Mehta}). The two-point correlation function is then a function of the distance $E_i-E_j$ and goes to zero for $E_i=E_j$. Such a correlation is due to the highly nontrivial operation of diagonalizing $H$, a procedure not mimicked properly in the examples above.

We thus conclude that the energy level spacings of an integrable model behave as independent variables inasmuch as they satisfy Poissonian statistics.  However as we have attempted to demonstrate, the energy levels of a non-integrable model must
exhibit highly non-trivial correlations in order to give GOE statistics.

\section{Studying perturbed minimal models using the truncated conformal space approach }

In this section we discuss how the level spacing statistics of the perturbed 
minimal models can be studied using the truncated conformal spectrum approach.  
We begin by reviewing briefly some necessary details of the minimal models of conformal field theory.

\subsection{The conformal minimal models}

The minimal models are a sequence of two dimensional conformally 
invariant quantum field theories.  They are labelled, $M_{p,q}$, where
$p$ and $q$ are two coprime integers (a unitary model is obtained for q=p+1).  
They have central charge of less than 1 given by 
\begin{equation}
c = 1-6\frac{(p-q)^2}{pq}.
\end{equation}
The field content of $M_{p,q}$ consists of
a set of primary fields, $\phi_{r,s},\bar{\phi}_{r,s}$ with chiral dimensions
\begin{equation}
h_{r,s} = \bar{h}_{r,s} = \frac{(pr-qs)^2-(p-q)^2}{4pq}, ~~~ 1\leq r < q, ~~  1\leq s < p.
\end{equation}
The Hilbert space of these theories consists of highest weight states 
\[
|h_{r,s},\bar{h}_{r,s}\rangle\equiv \phi_{r,s}(0)\bar{\phi}_{r,s}(0)|0\rangle,
\] 
(where we take $h=\bar{h}$ so restricting ourselves to the diagonal theories), acted upon by combinations of Virasoro generators, $L_n$,$\bar{L}_n$, satisfying
the algebra
\begin{equation}
[L_n,L_m]=(n-m)L_{n+m}+\delta_{n+m,0}\frac{c}{12}n(n^2-1).
\end{equation}
The set of states obtained by acting upon a single highest weight state with the Virasoro generators is known as a Verma module. On a cylinder of circumference, $R$, where time is taken along the cylinder's axis, the Hamiltonian and momentum are given by 
\begin{equation}
H_{{p,q}} \,=\, \frac{2\pi}{R}(L_0+\bar{L}_0-\frac{c}{12})
\,\,\,\,\,, \,\,\,\,\,\,\,
P_{p,q} \,=\, \frac{2\pi}{R}(L_0-\bar{L}_0).
\end{equation}
(we denote the sequence of unitary models by $H_p$ and $P_p$). 
This set of Hamiltonians has an infinite set of conserved charges which can be written
as normal ordered products of polynomials of the stress energy tensor \cite{Baz}.

We study deformations of the minimal models arrived at by perturbing them by some combination of fields in the theory:
\begin{equation}\label{per_mm}
H_{{p,q}} + \sum_i g_i \int^R_0 dx \,\phi_{r_i,s_i}(x).
\end{equation} 
These perturbations in general destroy the infinite set of conserved quantities present at the conformal point. The only deformations which generically give rise to integrable theories away from criticality are perturbations involving one (and only one) 
of the fields $\phi_{12}$, $\phi_{13}$ and $\phi_{21}$ \cite{Zamo1}. 
Perturbations involving multiple fields generically lead to non-integrable Hamiltonians \cite{Mussardo,MT}. 

\subsection{Minimal form of TCSA}

To study the Hamiltonians in Eqn.(\ref{per_mm}) we employ the truncated spectrum approach.  Because we have complete control over the conformal Hilbert space, we can enumerate all the states in the Hilbert space.  Although the conformal level spacing is finite (and essentially equal to $2\pi/R$), the continuum nature of the theory means the Hilbert space is nonetheless infinite dimensional.  In order to make the theory tractable we truncate the spectrum, keeping states whose energy in the conformal field theory is less than some cutoff, $E_c$.  Being left with a finite Hilbert space, we can form the Hamiltonian matrix through the evaluation of the following set of matrix elements:
\begin{equation}
\langle \Psi|H|\Psi' \rangle =
\frac{2\pi}{R}(\Delta_\Psi+\bar{\Delta}_\Psi-\frac{c}{12})\delta_{\Psi,\Psi '}
+ \sum_i g_i R\langle \Psi |\phi_{r_i,s_i}(0)|\Psi ' \rangle\delta{s_\Psi, s_{\Psi'}} ,
\end{equation}
where $|\Psi\rangle$ and $|\Psi'\rangle$ are two typical states in the truncated spectrum.
Here $L_0|\Psi\rangle = \Delta_\Psi|\Psi\rangle$ and we suppose the state $|\Psi\rangle$ has spin s.  The matrix elements $\langle \Psi |\phi_{r_i,s_i}(0)|\Psi ' \rangle$ are given in terms of the structure constants $C_{\Psi,\Psi'}^{h_{r_i,s_i}}$ of the theory and the chiral dimension of the perturbing operator:
\begin{eqnarray}
\langle \Psi |\phi_{r_i,s_i}(0)|\Psi ' \rangle
= \frac{(2\pi)^{2h_{r_i,s_i}-1}}{R^{2h_{r_i,s_i}-2}}C_{\Psi,\Psi'}^{h_{r_i,s_i}}M_{\Psi,\Psi'},
\end{eqnarray}
where $M_{\Psi,\Psi'}$ is some number depending on the particular Virasoro modes acting on the highest weight state in creating the state $|\Psi\rangle ,|\Psi '\rangle$. The structure constants of minimal models were computed by Dotsenko and Fateev 
\cite{DotFat}. 
Once these matrix elements are evaluated, we can easily diagonalize numerically the corresponding matrix and thus extract both the spectrum and matrix elements of the full (truncated) theory. 

For large values of the truncation energy, $E_c$, computing this matrix can involve the computation of millions of matrix elements.  As such, this computation has been automated.  To reduce the numerical burden further we perform the computations in the independent subspaces of the Hamiltonian (typically a given perturbation of a minimal model will not mix together all the Verma modules).  We also perform
the determination of the independent subspaces automatically through the computation of the structure constants, $C_{\Psi,\Psi'}^{h_{r_i,s_i}}$.  All told, the code we have developed to study the level spacings in perturbed minimal models can do so for arbitrary $M_{p,q}$ and for arbitrary perturbations.

\subsection{NRG improvements on TCSA}

In performing this computation upon the truncated Hamiltonian, the larger we choose the cutoff, $E_c$, the better.  However increasing $E_c$ comes with a numerical cost.  The number of states in the truncated theory grows exponentially with $E_c$. For a given Verma module, the number of states will grow as roughly the number, $P(N)$, of partitions of the integer $N=RE_c/2\pi$:
\begin{equation}\label{Hardy}
P(N) \simeq \frac{1}{4 N \sqrt{3}} e^{\pi \sqrt{2N/3}}, ~~~ N\rightarrow\infty .
\end{equation}
For the minimal CFTs, this growth is slightly slower due to the 
presence of zero norm states, but nonetheless is still exponential.

A way in which to improve upon the precision with which the high energy levels are handled without exponentially increasing the computational burden is given by a numerical renormalization group (NRG) approach.  This approach, modelled after Wilson's numerical renormalization group for quantum impurities, allows one to keep hundreds of thousands of states in the truncated Hilbert states (as opposed to order ten thousand without NRG \cite{Konik}).

The basic idea behind the approach is the ability to order the states of the CFT
in terms of their importance for determining the (relatively) low energy behaviour of the full theory.  This ordering is determined by the unperturbed energies of the states of the CFT field theory.  Because the perturbation of the CFT that we are considering 
is relevant, low energy states of the CFT are more important for determining the low
energy physics than high energy states are.  The situation is analogous to Wilson's
treatment of the Kondo problem.  There Wilson mapped a Kondo impurity interacting with free electrons onto a tight binding model of non-interacting electrons on a lattice extending from 0 to $\infty$ interacting with an impurity located at $x=0$.  Wilson reasoned that a site on the lattice close to the impurity will influence the physics more than a site further away.  Thus the natural way to proceed in such a problem is to first diagonalize the interactions of the Kondo impurity with the electrons on nearby lattice sites, and then take into account in a progressive fashion, sites further and further away from the impurity.  We mimic this procedure here.  However instead of distance
from the Kondo impurity as a metric of importance, we use the magnitude of the unperturbed energy of a conformal state.

The NRG algorithm by which a truncated perturbed conformal field theory is treated is as follows. We imagine ordering in energy the states of the unperturbed conformal field theory, $|0\rangle,|1\rangle, \cdots $. As a first step we take the $N+\Delta$ of the states lowest in energy.  We form the full Hamiltonian (both CFT and perturbation) in this truncated basis of $N+\Delta$ states. Just as with the ordinary TCSA, we diagonalize (numerically) the problem leaving us with $N+\Delta$ eigenstates, 
$E^1_1,\cdots ,E^1_{N+\Delta}$ and eigenvectors, $|E\rangle^1_1,\cdots ,|E\rangle^1_{N+\Delta}$.  
In the next step of the NRG algorithm, we order these $N+\Delta$ states in ascending 
order of their energies and toss away the top $\Delta$ states.  We express these remaining eigenvectors in terms of the conformal basis as follows:
\begin{equation}
|E\rangle^1_k = A^1_{km}|m\rangle + B^1_{km'}|m'+N\rangle; ~~~~k=1,\cdots,N,
\end{equation}
where $A^1$ is an $N\times N$ matrix and $B^1$ is an $N\times \Delta$ matrix and repeated indices are summed on.
To these $N$ eigenstates we add the next $\Delta$ states from the unperturbed theory,$|m+\Delta+1\rangle,\cdots,|m+2\Delta\rangle$.  
This leaves us again with a truncated Hilbert space of $N + \Delta$ states.  We reform the Hamiltonian in this new basis and then rediagonalize so extracting a new set of $N+\Delta$ energies, $E^2_1,\cdots ,E^2_{N+\Delta}$, and eigenstates, $|E\rangle^2_1,\cdots ,|E\rangle^2_{N+\Delta}$.  We again order the eigenstates in energy
and toss away the top most $\Delta$ states.  If we reexpress the remaining eigenstates, $|E\rangle^2_k$ in terms of the original conformal basis we obtain
\begin{eqnarray}
|E\rangle^2_k &=& A^2_{km}|E\rangle^1_m + B^2_{km}|m+N+\Delta\rangle\cr\cr
&=& (A^2A^1)_{km}|m\rangle + (A^2B^1)_{km'}|m'+N\rangle+ B^2_{km'}|m'+N+\Delta\rangle,
\end{eqnarray}
where again $A^2$ is an $N\times N$ matrix and $B^2$ is an $N\times \Delta$ matrix.  Here the index m runs from $1$ to $N$ while the index m' runs from $1$ to $\Delta$.

We can repeat this procedure {\em ad libitum}: we order the new set of $N+\Delta$ states, toss away the top most $\Delta$ states, add the next $\Delta$ set of unperturbed
states from the CFT, reform the Hamiltonian, rediagonalize, etc.
In this way, we allow the higher energy states of the unperturbed CFT to mix in with eigenstates of the full (but truncated) theory.  As we keep $N+\Delta$ fixed at each step, the associated computational problem grows only as the square of the total number of states kept (owing to the need to manipulate states (albeit only $N+\Delta$ of them) which are expressed in terms of an ever growing basis as the NRG proceeds).  

At the nth-iteration, the eigenstates have the form
\begin{eqnarray}
|E\rangle^n_k &=& (A^n\cdots A^1)_{km}|m\rangle + (A^n\cdots A^2B^1)_{km'}|m+N\rangle \cr\cr
&& \hskip -.5in +
(A^n\cdots A^3 B^2)_{km'}|m'+N+\Delta\rangle + \cdots \cr\cr 
&& \hskip -.5in + (A^n B^{n-1})_{km'}|m'+N+(n-2)\Delta\rangle
+ B^n_{km'}|m'+N+(n-1)\Delta\rangle.
\end{eqnarray}
We see that each term in the above sum has a matrix product state form.  The approximation encoded in the NRG
is then one where we study a Hamiltonian arrived at by projecting the original Hamiltonian onto a space
composed of matrix product states of the above form. 

The NRG scheme as described above is able to obtain accurately the low energy portion of the spectrum.  However as we interested in determine the level spacing distribution we need the first several thousand levels to high accuracy.  In order to obtain such accuracy we adopt a sweeping procedure not dissimilar to the finite volume
algorithm in DMRG \cite{White}.  Or equally good, this procedure may be considered as an extended set of Jacobi transformations done to diagonalize a symmetric matrix (extended in the sense that we are zeroing blocks not individual elements
of a matrix).  The procedure works as follows.  Suppose we have completed the NRG procedure as described above for $M$ iterations.  And in doing so we would have run through the first $N+M\Delta$ states of the conformal basis.  We now begin anew.  But instead of working with $N+M\Delta$ conformal states, we work with a basis formed from the eigenstates generated in the the NRG procedure.  This basis is given by
\begin{eqnarray}
&&|E\rangle^M_1,\ldots,|E\rangle^M_N,|E\rangle^1_{N+1},\ldots,|E\rangle^1_{N+\Delta},|E\rangle^2_{N+1},\ldots,
|E\rangle^2_{N+\Delta},\ldots\cr\cr
&& \hskip 1in |E\rangle^{M-1}_{N+1},\ldots,|E\rangle^{M-1}_{N+\Delta},|E\rangle^M_{N+1},\ldots,|E\rangle^M_{N+\Delta}.
\end{eqnarray}
We see that this basis is formed from taking the first $N$ states coming from the last NRG iteration (and so the best guess we have at the $N$ lowest energy states in the theory), followed by the $M\times\Delta$ states we discarded in the $M$ iterations of the NRG.  This basis is much closer to the true eigenstates of the system than the initial conformal basis was. Using this basis we repeat the above described set of NRG iterations (although because this basis is not orthonormal an additional step at each iteration is needed to form an orthonormal basis). This set of iterations constitutes a single sweep.  We find that the eigenenergies converge rapidly after only a handful
of sweeps.  To test this procedure we compared exact diagonalization of large matrices ($12000\times 12000$) with the NRG plus three sweeps with $N=4000$ and $\Delta=1000$.  We found that this number of sweeps was enough to guarantee that 90\% of the first N eigenstates matched the exact diagonalization to 4 significant digits.

\section{Level Spacing Results}

The results of any study of the level spacings of a perturbed minimal model will depend crucially on the value of the system size, $R$.  For a generic perturbation, the model will be massive with some mass scale $m=\xi^{-1}$. This allows us to identify two different regimes: $R \ll \xi$  and $R \gg \xi$.  In the regime $R \ll \xi$, the theory is in its UV limit.  Regardless of whether the perturbation is integrable or otherwise, the level spacing distribution will be close to being Poissonian. On the other hand for $R \gg \xi$, the theory is in its IR limit and whether the theory is integrable or non-integrable becomes important for determining the distribution its level spacing obeys.  A consequence of this is that as $R$ is varied for a non-integrable model the level spacing distribution will evolve from Poissonian to GOE.
 
\begin{figure}[ht]
\begin{center}$
\begin{array}{cc}
\text{(a)}\includegraphics[width=0.40\textwidth]{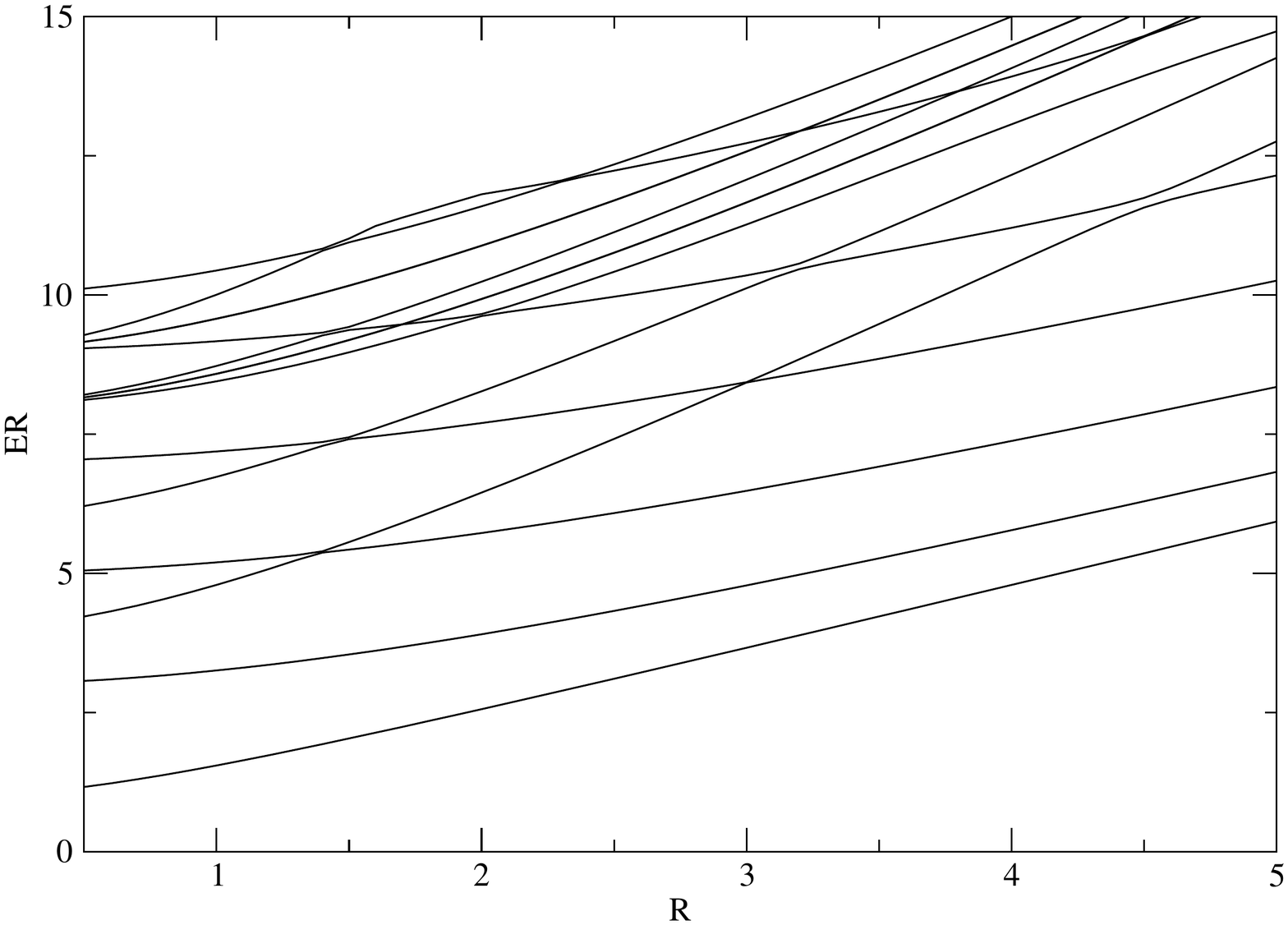} 
&\text{(b)} \includegraphics[width=0.40\textwidth]{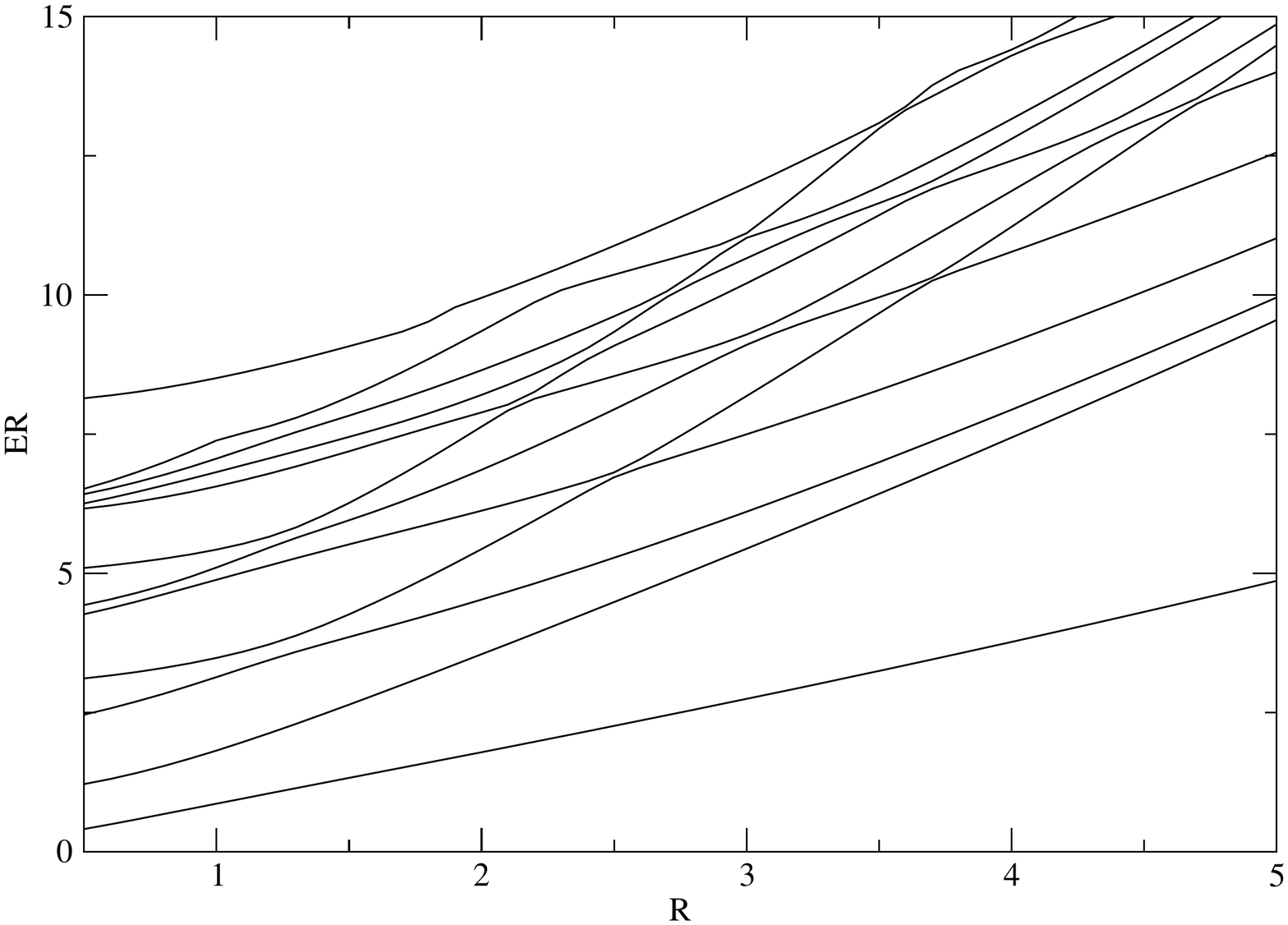} 
\end{array}$
\end{center}
\label{int-nonint}
\caption{Energy levels(multiplied by R) vs cylinder size. (a) Integrable case, showing several level crossings. (b) Non-integrable case, showing instead avoided crossings.}  
\end{figure}

The truncation energy, $E_c$ used in studying a given theory with the TCSA also influences the level spacing distribution.  To see this first imagine that we are in the IR limit and we plot the energy levels as a function of R for the integrable and non-integrable case.  Further imagine that the truncation is taken to infinity (there is no truncation of the spectrum).  In the integrable case, we see that energy levels cross at various values of R (schematically pictured in Figure 4.1a).  This occurs because even though we work in a given symmetry sector where all the quantum numbers of the trivial symmetries are fixed, the eigenstates still carry the quantum numbers of the remaining non-trivial symmetries that are present in the integrable model.  If these quantum numbers are different, the eigenstates cannot mix and their corresponding energies may cross. This crossing leads to weight in the level spacing distribution for arbitrarily small level spacings.  While it does not explain the appearance of a Poissonian distribution, it is at least consistent with it. If however integrability is not present, eigenstates will be able to mix generically and what we will see instead is a set of avoided crossings (see Figure 4.1b).

Now imagine that we include the effects of $E_c$.  The presence of a finite truncation of an integrable theory will generically destroy integrability.  (The infinite set of conserved charges in the theory, being polynomials of the stress-energy tensor, involve Virasoro modes of arbitrary high grade.  But in the presence of a finite $E_c$, modes with
grade greater than $RE_c/2\pi$ are discarded from the theory so breaking the charges' structure.)  Correspondingly level crossings will turn into avoided crossings.  Thus a theory with a finite $E_c$ should not have a perfect Poissonian distribution.  But of course if $E_c$ is chosen to be large enough, the breaking of integrability
will be sufficiently weak that the distribution will remain nearly
Poissonian.  One reason it is necessary to use the numerical renormalization group in this case is to be able to set $E_c$ high enough so that this breaking is in fact small.

The effects of $E_c$ are not independent of the system size, $R$, the relevant dimensionless parameter being $\xi E_c/2\pi = \xi 2 n_c /R$, where $n_c$ is the level at which we truncate the chiral Verma modules.  Thus for even a large value of $E_c$, a sufficiently large value of $R$ will break integrability.  Thus for an integrable model, as we vary $R$ we expect to see the level spacing distribution to evolve from Poissonian to something GOE-like.  In the integrable case, we will be particularly interested in values of $R$ large enough such the theory is in its massive phase (i.e. $R > \xi$) but where $\xi E_c/2\pi$ is sufficiently large that integrability breaking is weak.

To quantitatively characterize a given distribution $g(s)$, it is common (see \cite{eta} for example) to define a quantity $\eta$:
\be
\eta=\frac{\int_0^\infty ds |g(s)-e^{-s} | }{\int_0^\infty ds | \frac{\pi s}{2} e^{-\frac{\pi}{4} s^2}- e^{-s} |} .
\ee
Note that our definition of $\eta$ is differs slightly from the common one. Usually the integration is restricted in the region $0\leq s\leq s_0$, where $s_0$ is the first intersection point between Poissonian and GOE. Such a definition completely disregards the tail of the distribution. However the tail is less sensitive to the bin size chosen, and has very different behavior in the Poissonian and GOE case. Therefore, disregarding the tail gives a much less reliable estimate for $\eta$.  

From the above definition, we can extract the limiting behavior of $\eta$.
If $\eta$ is close to 1, $g(s)$ is GOE-like.  If instead it is close to 0, $g(s)$ is Poissonian.  
And finally if it is close to $\frac{1}{2}$, the distribution $g(s)$ is similar to (Poisson+GOE)/2.

\subsection{Tricritical Ising model}
We now consider as a first example the level spacing statistics of the perturbed tricritical
Ising model\footnote{We have also performed similar calculations, with similar results, for the usual Ising model, but we prefer to focus our discussion here on the TIM for its larger number of integrable and non-integrable deformations.} (for previous TCSA studies of this model see \cite{LMT,LMC}). The tricritical Ising model (TIM) is described by the minimal model $M_{4,5}$. The field content is characterized by six primary fields. We will consider the perturbations by the fields $\phi_{1,2}$ (the leading energy perturbation) and $\phi_{2,2}$ (the leading spin perturbation) and a combination thereof. The scaling dimensions of these two fields are respectively $\frac{1}{10}$ and $\frac{3}{80}$. The perturbation $\phi_{2,2}$ is odd under $\mathbb{Z}_2$ symmetry while $\phi_{1,2}$ is even. In analyzing these theories, we focus on the zero momentum subsector.  In the cases where $\sigma \rightarrow -\sigma$ is a good symmetry ($\sigma$ is the spin field) we analyze the largest subspace.

The first results we present are level spacings obtained for the integrable perturbation $\phi_{1,2}$ of the TIM. The coupling constant is set to $0.092834\ldots$, to ensure a mass gap, $m=1$ \cite{Fateev}. For a truncation at level $n_c=14$, the zero momentum 
Hilbert space splits in two subspaces of 22559 and 18751 states.  We employ the NRG with a base matrix of $N=4000$, a step size of $\Delta=1000$, and use three sweeps to obtain convergence. Of the 4000 eigenvalues in the base matrix, we employ the lowest 3500 in computing the level spacing distribution. For smaller values of $R$ we are in the conformal limit.  
\begin{figure}[ht]
\begin{center}$
\begin{array}{ccc}
\text{(a)}\includegraphics[width=0.30\textwidth]{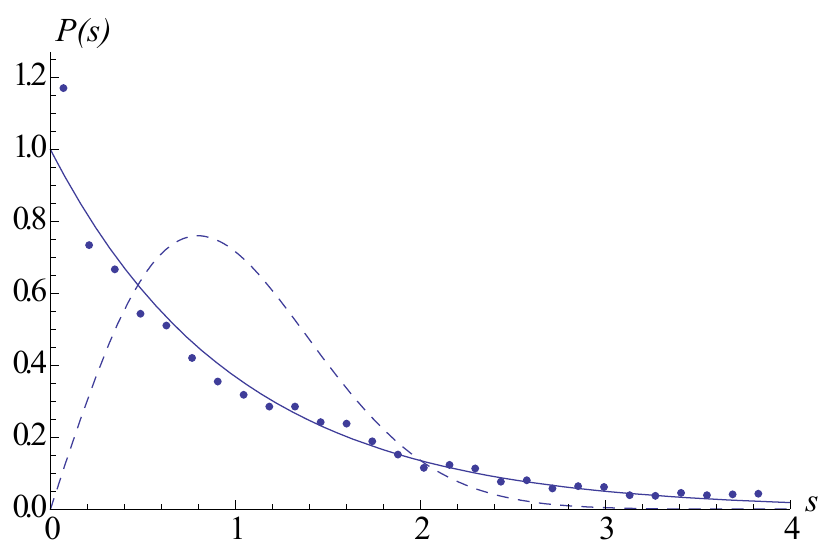} 
&\text{(b)} \includegraphics[width=0.30\textwidth]{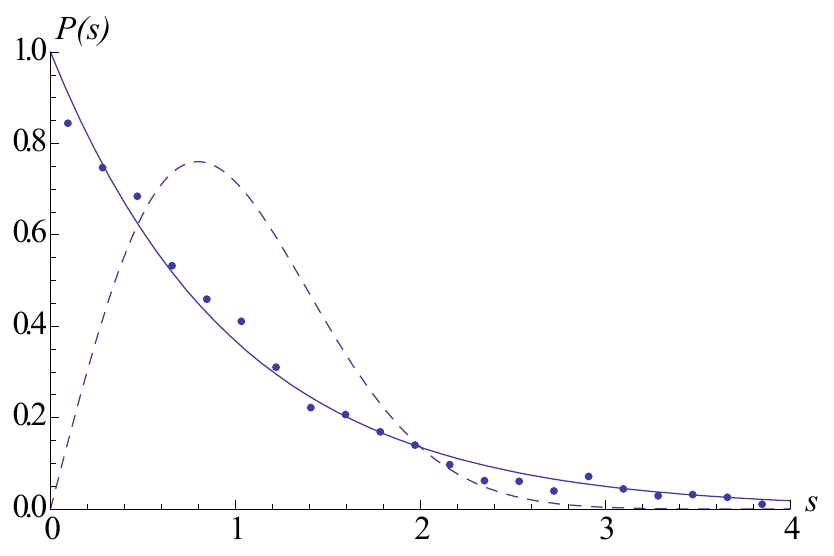} 
&\text{(c)}\includegraphics[width=0.30\textwidth]{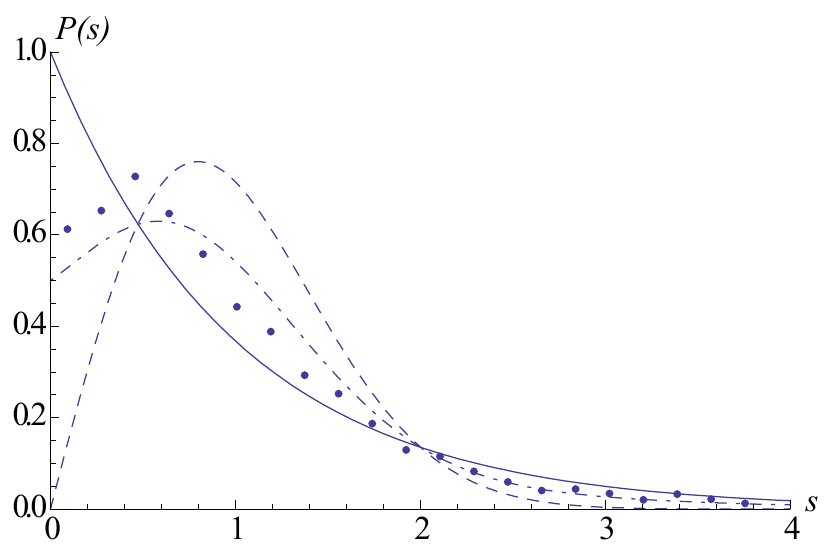} 
\end{array}$
\end{center}
\label{M45-12}
\caption{Level spacing distribution for $M_{4,5}+\phi_{1,2}$ for three values of $mR$: (a) mR=18.9, 
$\eta$=0.25, correlation coefficient for the fit to a Poissonian distribution, $r_P=0.98$; 
(b) mR=22, $\eta$=0.22, $r_P=0.99$ (c) mR=25.2, $\eta$=0.55, $r_P=0.95$, correlation coefficient for the fit
to (Poissonian+GOE)/2, $r_{AV}=0.98$.  In these three plots we also display a pure Poissonian
distribution (solid line) and a pure GOE distribution (dashed line).}  
\end{figure}
 
\begin{figure}[ht]	
\begin{center}$
\begin{array}{ccc}
\text{(a)}\includegraphics[width=0.30\textwidth]{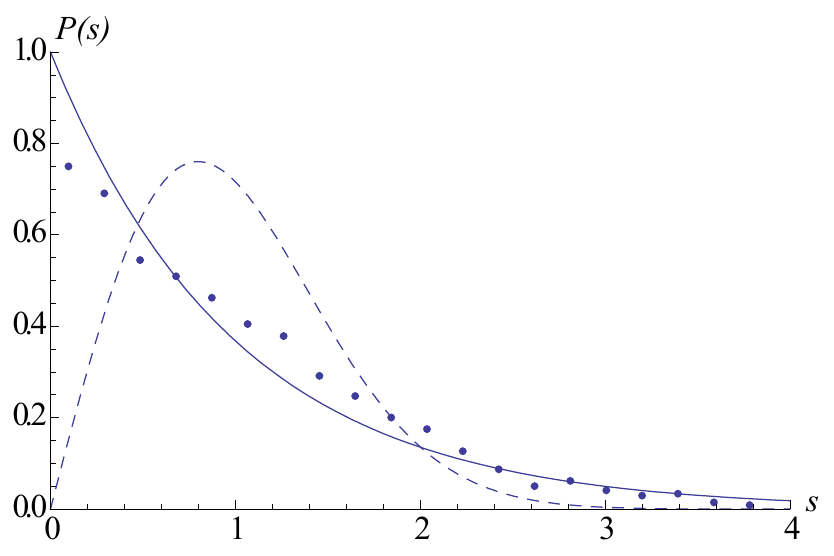} & 
\text{(b)}\includegraphics[width=0.30\textwidth]{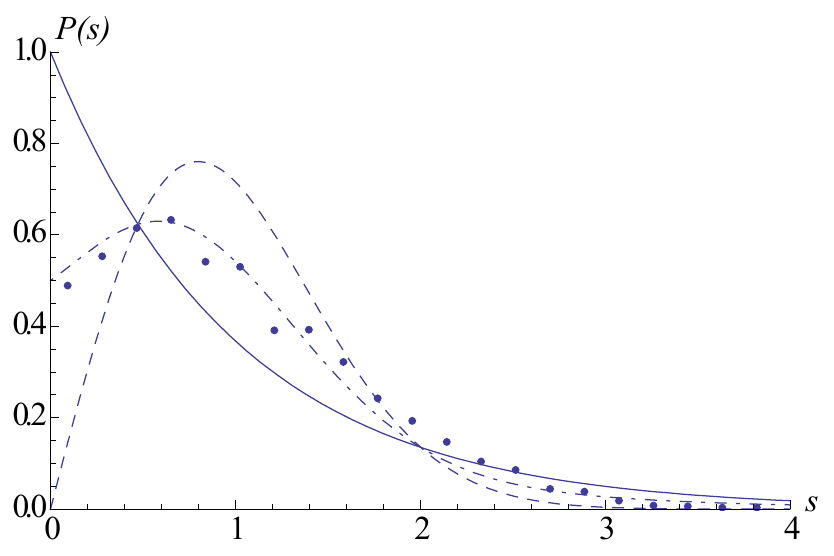} &
\text{(c)}\includegraphics[width=0.30\textwidth]{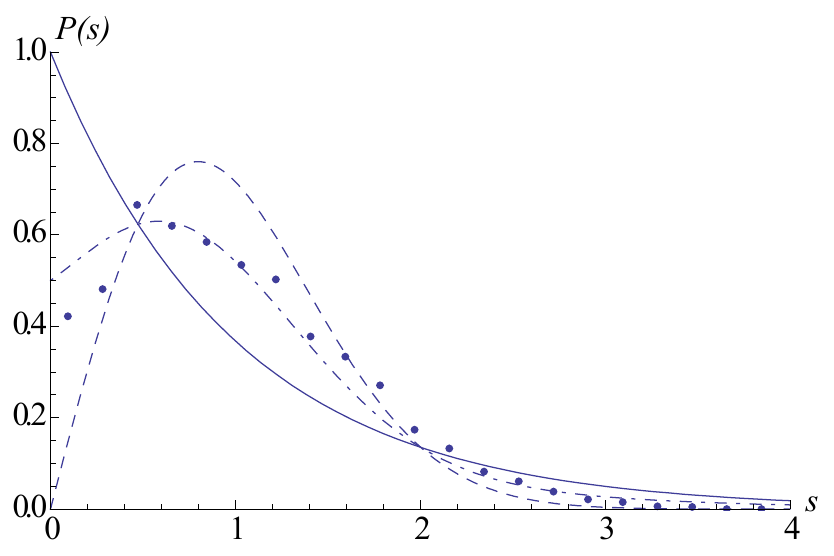} 
\end{array}$
\end{center}
\begin{center}$
\begin{array}{cc}
\text{(d)}\includegraphics[width=0.30\textwidth]{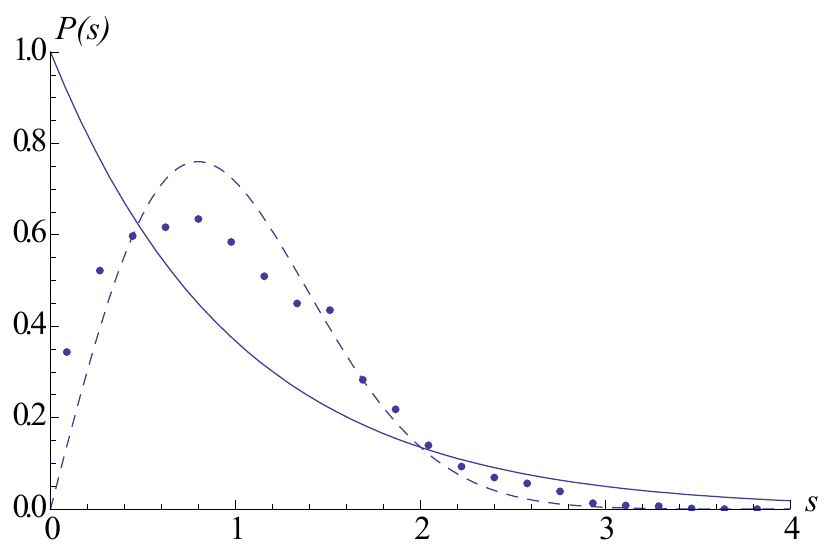} &
\text{(e)}\includegraphics[width=0.30\textwidth]{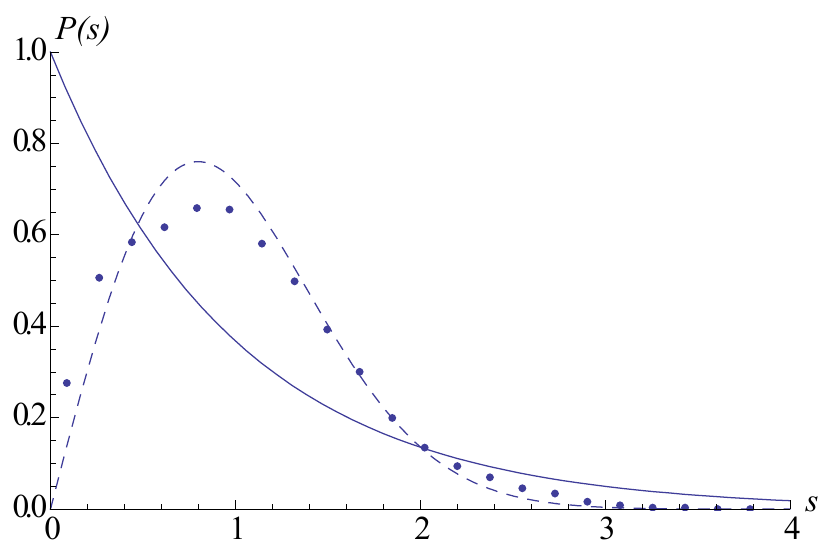} 
\end{array}$
\end{center}
\caption{Level spacing distributions for $M_{4,5}+\phi_{2,2}$ (a) mR=8.2, $\eta =0.25$, $r_P=0.98$; 
(b) mR=13.2, $\eta=0.52$, $r_{GOE}=0.91$ (the correlation coefficient w.r.t. the GOE distribution), $r_{AV}=0.99$;
(c) mR=18.9, $\eta=0.64$, $r_{GOE}=0.95$, $r_{AV}=0.98$; 
(d) mR=22, $\eta=0.68$, $r_{GOE}=0.97$, $r_{AV}=0.98$; 
(e) mR=25.2, $\eta=0.75$, $r_{GOE}=0.98$.}
\label{M45-22}
\end{figure}
In the pure conformal limit, all energy levels are of the form $\frac{2\pi}{R}(n-1/12)$ where n is an integer. Moreover, the degeneracy of each level n grows exponentially in n.  Thus we might expect a disproportionate weight at zero in the energy level spacings.  This is what we find in Figure 4.2a.

As we increase the value of $R$ we see that we obtain a level distribution that is Poissonian at all values of $s$ (see Figure 4.2b).  As we increase $R$ further we see that the Poissonian distribution evolves into a distribution closer to a mix of Poissonian and GOE (Figure 4.2c).  As we have argued, this is to be expected as for larger values of $R$, we expect the integrability of the theory to be compromised by the presence of a finite energy cutoff.

\begin{figure}[ht]
\begin{center}$
\begin{array}{cc}
\text{(a)}
\includegraphics[width=0.30\textwidth]{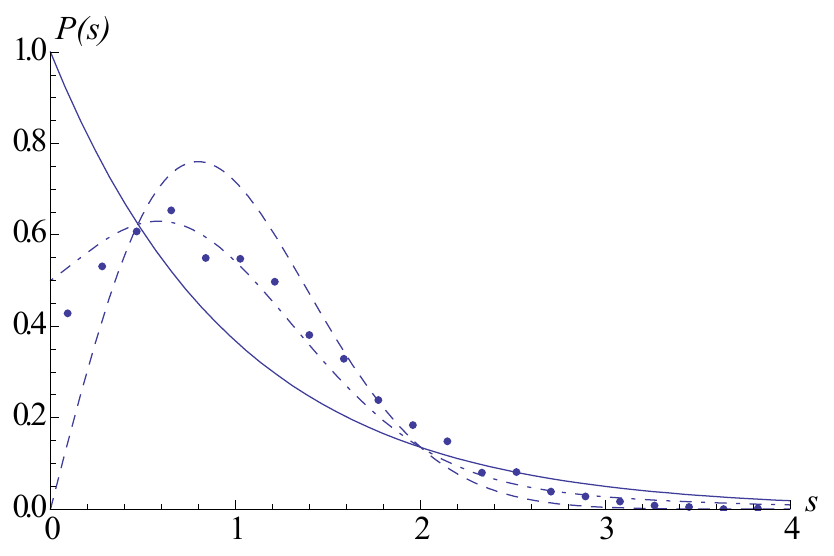} & 
\text{(b)}\includegraphics[width=0.30\textwidth]{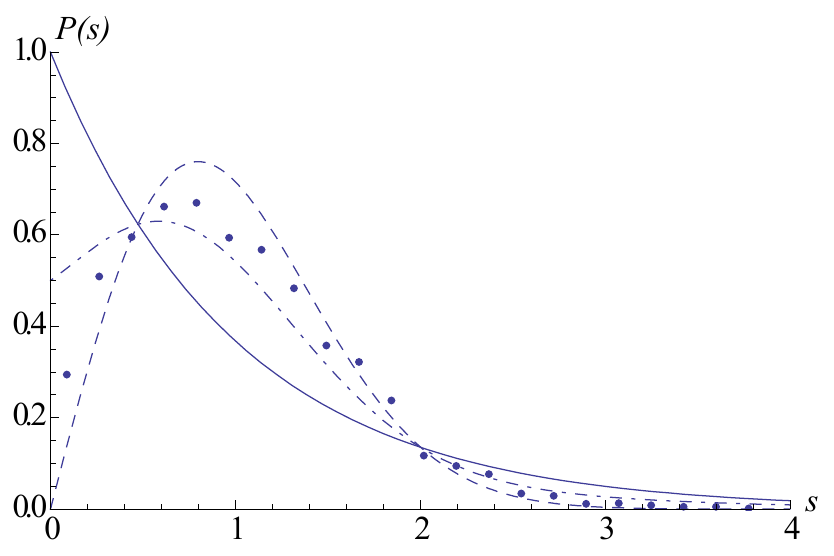} 
\end{array}$
\end{center}
\begin{center}$
\begin{array}{cc}
\text{(c)}\includegraphics[width=0.30\textwidth]{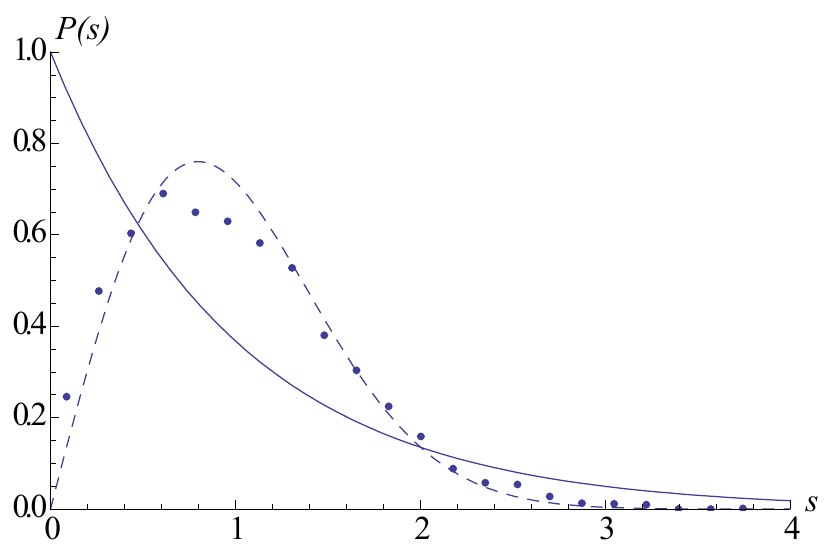} &
\text{(d)}\includegraphics[width=0.30\textwidth]{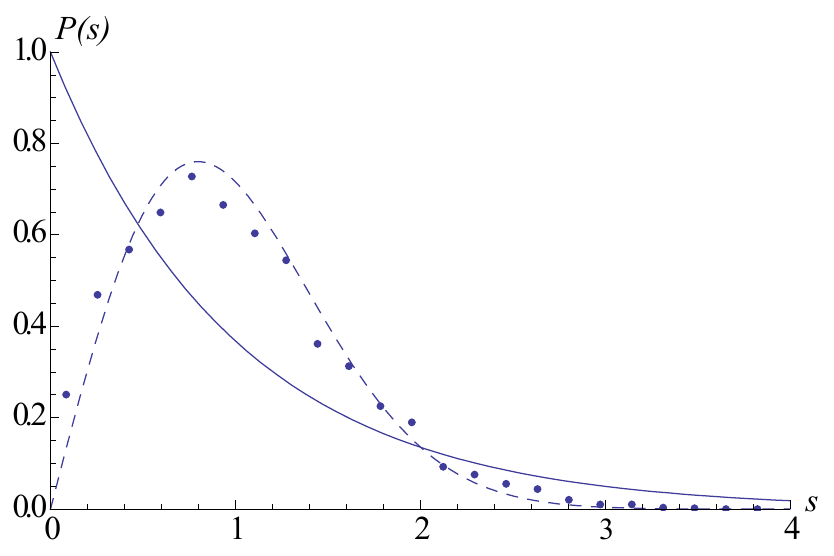} 
\end{array}$
\end{center}
\caption{Level spacing distribution for $M_{4,5}+\phi_{1,2}+\phi_{2,2}$
(a) mR=13.2, $\eta=0.60$, $r_{GOE}=0.94$, $r_{AV}=0.99$; 
(b) mR=18.9, $\eta=0.74$, $r_{GOE}=0.98$;
(c) mR=22, $\eta=0.79$, $r_{GOE}=0.99$;
(d) mR=25.2 $\eta=0.80$, $r_{GOE}=0.99$.}
\label{M45-12-22}
\end{figure}

For the next example, we consider the non-integrable $\phi_{2,2}$ perturbation of the TIM. Again we chose the coupling constant so that $m=1$. The truncated Hilbert space consists of 41310 states. The NRG base matrix $N$, step $\Delta$ and number of states considered are chosen as in the $\phi_{1,2}$ case. The level spacing distributions for various values of $mR$ are shown in Figure 4.3.
For very small values of $mR$ we see that the distribution is Poissonian as expected.  As $mR$ is increased the distribution becomes more and more GOE-like.  
We may ask whether this effect is aided by truncation effects.  We see to some degree that it is. At the values of $mR$ shown, the distribution is most GOE like for the largest value of $mR$ (25.2). But from our study of $M_{4,5}$+$\phi_{1,2}$ we know that at this value of $mR$ the Poissonian distribution due to the model's integrability has already begun to break down.

As a final case for the TIM, we consider perturbing the theory by both $\phi_{1,2}$ and $\phi_{2,2}$ simultaneously. 
The coupling constants are chosen such that $m=1$.
The Hilbert space and $N$ and $\Delta$ have the same size as the $\phi_{2,2}$ case.  The results
are presented in Figure 4.4.
As with $M_{4,5}+\phi_{2,2}$, we see that for $M_{4,5}+\phi_{1,2}+\phi_{2,2}$ increasing the value of $R$ leads to a distribution
that is more and more GOE like.

\subsection{Tetracritical Ising Model}
To demonstrate that our level spacing results possess a certain universality, we
consider the tetracritical Ising model, a theory
described by the minimal model $M_{5,6}$. (Its partition function differs from three state
Potts in that it is diagonal in the Virasoro characters.) The field content of this theory 
is characterized by ten primary fields. We will consider, separately, the perturbations $\phi_{1,2}$ (integrable)
and $\phi_{2,2}$ (non-integrable). 
The scaling dimensions of these fields are respectively $\frac{1}{8}$ and $\frac{1}{40}$. 

\begin{figure}[ht]
\begin{center}$
\begin{array}{cc}
\text{(a)}\includegraphics[width=0.30\textwidth]{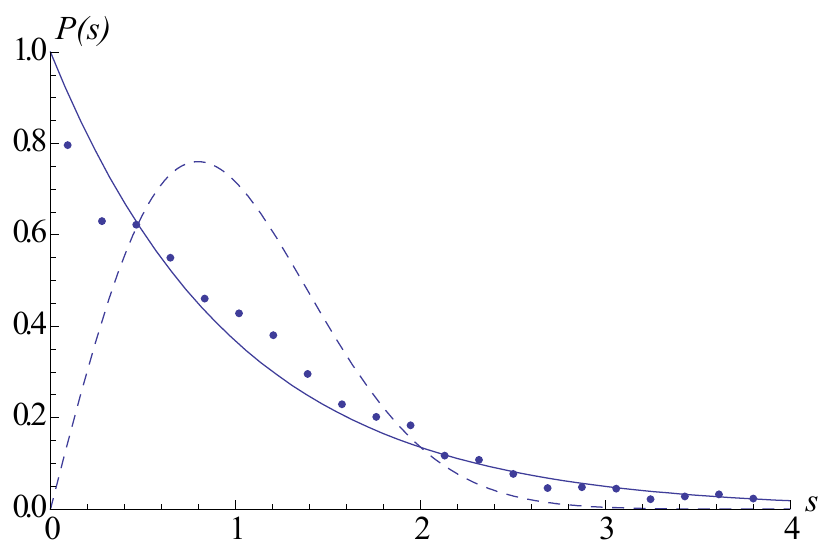} & 
\text{(b)}\includegraphics[width=0.30\textwidth]{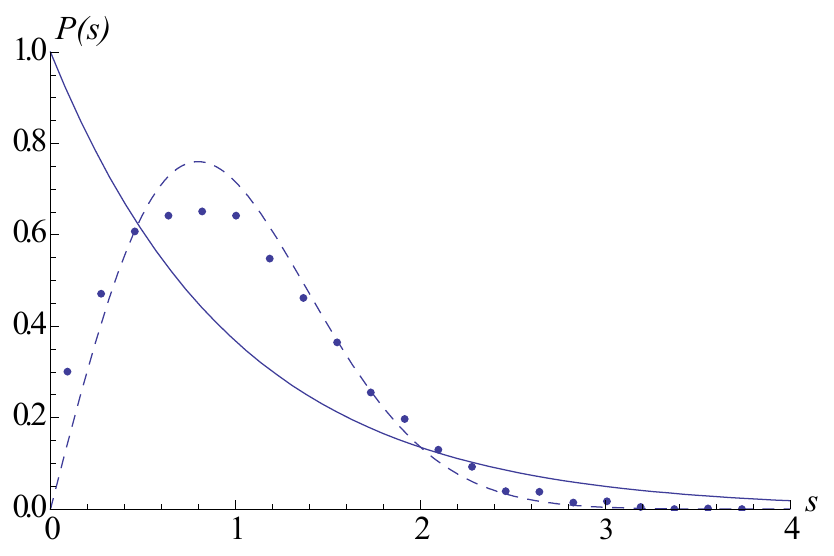} 
\end{array}$
\end{center}
\caption{(a) Level spacing distribution for $M_{5,6}+\phi_{1,2}$;
mR=18.9, $\eta=0.20$, $r_P=0.98$; 
(b) Level spacing distribution for $M_{5,6}+\phi_{1,2}$;
mR=18.9, $\eta=0.75$, $r_{GOE}=0.98$.}
\label{M56-12-22}
\end{figure}

We take the truncation level to be $n_c=11$ with the corresponding size of the truncated Hilbert space being 27931 states for $\phi_{2,2}$  case and 17601 for $\phi_{1,2}$.
We set the coupling constants in both cases such that the mass gap is again 1. 
We see in Figure \ref{M56-12-22} that, as expected, the integrable perturbation
has Poissonian level spacing statistics while the non-integrable perturbation has GOE statistics.

\section{Conclusions}

In this work we have studied the level spacing statistics of perturbations of the conformal minimal models.  We find that the level spacing statistics of the integrable variants of such models are Poissonian while their non-integrable counterparts obey GOE statistics.  This general result does come with two caveats.  For values of the system size, $R$, of up to 10 times the correlation length, $\xi$, the system exhibits Poissonian-like statistics regardless of the whether the perturbation is integrable or not, a result of being in the conformal limit (at least at higher energies) for moderate values of $R$.  The second caveat is that at very large values of $R$, values where $\xi E_c$ is less than 7, the level spacing distribution becomes GOE-like regardless of the presence of integrability.  This results from the presence of a finite energy cutoff in our numerical treatment of these theories.  At large values of $R$ (or equivalently small values of the dimensionless parameter $E_c\xi=\frac{4\pi n_c\xi}{R}$), the presence of this truncation leads to a strong breaking of any symmetries in the model (and so integrability), so guaranteeing GOE like statistics.

To determine the level spacing statistics we employed the truncated conformal spectrum approach.  In this method the underlying conformal spectrum is truncated at some energy $E_c$. In order to obtain results that approximate the theory without cutoff, we needed to chose $E_c$ to be large enough that the considered Hilbert space contained in certain cases on the order of 40000 states. This cannot be handled directly by exact diagonalization.  Thus we used a numerical renormalization group procedure \cite{Konik} which enables large truncated Hilbert spaces to be studied in a numerically feasible fashion.  While this approach yields accurate results for low lying eigenstates, it, by itself, cannot guarantee accuracy over the thousands of levels needed to determine the level spacing distribution.  Thus we developed a sweeping procedure guaranteeing convergence of high energy eigenstates to their true value.

In this paper we report the results of our analysis for two particular minimal models, the tricritical Ising model and the tetracritical Ising model. These results point to the infinite set of quantum Hamiltonians associated to the deformations of conformal minimal models as an ideal playground for studying important characteristics of generic non-integrable quantum systems. It would be interesting, for instance, to employ these models in order to test the validity of the various scenarios put forward for the thermalization of generic isolated quantum systems following a sudden quench \cite{Rigol}. Moreover, properly modified, the numerical method developed here can be also used to address the computation of non-equilibrium and time-dependent correlation functions of various order parameters. 

\vskip .5in
\noindent Acknowledgements: GB and GM acknowledge the grants INSTANS (from the ESF) and 2007JHLPEZ (from MIUR).  RMK acknowledges support by the US DOE under contract number DE-AC02-98 CH 10886.

\end{document}